\documentclass{jfmc}
\usepackage{graphicx}
\usepackage{subfig}
\usepackage{tikz}
\usepackage{adjustbox}
\usepackage{amsmath}
\usepackage{hyperref}

\DeclareMathOperator\erfc{erfc}

\shorttitle{Drag correlations for the volume-filtered framework}
\shortauthor{van Wachem, Elmestikawy, Chandran, and Hausmann}

\title{A new paradigm for computing hydrodynamic forces on particles in Euler-Lagrange point-particle simulations}

\author{Berend~van~Wachem\corresp{\email{berend.van.wachem@multiflow.org}}, 
  Hani~Elmestikawy, Akshay~Chandran,
 \and Max~Hausmann}

\affiliation{Chair of Mechanical Process Engineering, Otto-von-Guericke-Universität Magdeburg, Universitätsplatz 2, 39106 Magdeburg, Germany}

\begin{document}

\maketitle

\begin{abstract}
Accurate prediction of the hydrodynamic forces on particles is central to the fidelity of Euler-Lagrange (EL) simulations of particle-laden flows. Traditional EL methods typically rely on determining the hydrodynamic forces at the positions of the individual particles from the interpolated fluid velocity field, and feed these hydrodynamic forces back to the location of the particles. This approach can introduce significant errors in two-way coupled simulations, especially when the particle diameter is not much smaller than the computational grid spacing. In this study, we propose a novel force correlation framework that circumvents the need for undisturbed velocity estimation by leveraging volume-filtered quantities available directly from EL simulations. Through a rigorous analytical derivation in the Stokes regime and extensive particle-resolved direct numerical simulations (PR-DNS) at finite Reynolds numbers, we formulate force correlations that depend solely on the volume-filtered fluid velocity and local volume fraction, parametrized by the filter width. These correlations are shown to recover known drag laws in the appropriate asymptotic limits and exhibit a good agreement with analytical and high-fidelity numerical benchmarks for single particle cases, and, compared to existing correlations, an improved agreement for the drag force on particles in particle assemblies. The proposed framework significantly enhances the accuracy of hydrodynamic force predictions for both isolated particles and dense suspensions, without incurring the prohibitive computational costs associated with reconstructing undisturbed flow fields. This advancement lays the foundation for robust, scalable, and high-fidelity EL simulations of complex particulate flows across a wide range of industrial and environmental applications.
\end{abstract}

\begin{keywords}
Eulerian-Lagrangian point particle method; Particle-source-in-cell, Volume-filtered Navier-Stokes equations; Drag force modelling.
\end{keywords}

\section{Introduction}
\label{sec:introduction}

Particle-laden flows are prevalent in both natural phenomena and various industrial processes. Despite extensive research efforts, our comprehension of the collective behaviour of particles within a fluid flow remains incomplete. Alongside experimental studies, numerical modelling of particle-laden flows has gained substantial importance over the past few decades, driven by advancements in computational power and the development of more efficient and precise numerical methods.

The array of numerical methods available for simulating particle-laden flows spans from ``particle-resolved'' methods, such as the immersed-boundary method (IBM), which resolves the flow around each individual particle on a fine computational mesh~\citep[\textit{e.g.},][]{Peskin1972}, to fully interpenetrating Eulerian approaches, in which both the fluid and particle phases are treated as continuous media~\citep[\textit{e.g.},][]{Anderson1967}. Positioned between these two extremes is the Euler-Lagrange approach, which treats the fluid phase as a continuum, while resolving the trajectories of individual particles.

Euler-Lagrange (EL) point-particle methods are particularly effective for simulating fluid flows laden with up to several millions of particles, providing an accurate, straightforward, and cost-efficient solution. In EL approaches, the fluid phase dynamics are solved using a classical Eulerian framework, whereas the positions of the particles, treated as Lagrangian point-masses, are evolved based on the computed fluid flow field. The forces acting on the particles, such as drag, are typically estimated using semi-empirical models~\citep[\textit{e.g.,}][]{Schiller1933,Wen1966,Ergun1952}.

Different levels of coupling between the fluid and particulate phases can be considered, each suitable for different particle volume fraction and mass loading regimes. For very dilute particle-laden flows with very low particle volume fractions and/or mass loading, one-way coupling is often assumed. In this scenario, the momentum transfer from the particles to the fluid phase is negligible, meaning the flow is assumed to be unaffected by the presence of the particles. However, when the particle volume fraction exceeds approximately $10^{-5}$, the momentum transfer becomes significant and cannot be ignored~\citep{Sommerfeld2008}. In such cases, where two-way coupling is applied, the particles influence the flow through source terms in the governing fluid momentum equations.

From a computational perspective, the momentum transfer between the particles and the fluid in the Euler-Lagrange (EL) framework is commonly addressed using the particle source-in-cell (PSIC) model proposed by~\citet{Crowe1977}. This model has become a cornerstone in simulating particle-laden flows due to its ability to incorporate the interactions between discrete particles and the continuous fluid phase effectively and has been extensively utilized over the past decades to model particle-laden flows,~\citep[\textit{e.g.},][]{Marchioli2008,Eaton2009,vanWachem2001b}. However, one of the important assumptions of this approach is that the ratio $d_\mathrm{p}/\Delta x$ is very small, where $d_\mathrm{p}$ is the particle diameter, and $\Delta x$ represents the computational mesh spacing. In our recent work~\citep{Evrard2021}, we show that the error of the PSIC-EL method in the Stokes regime is linearly proportional to the ratio $d_\mathrm{p}/\Delta x$, and can be as large as 10\%, even if following the recommendations of 
commonly used best-practice guidelines~\citep{Sommerfeld2008}, which advises $d_\mathrm{p}/\Delta x < 0.1$. 
These errors have contributed to poor results in several detailed validation studies conducted using the PSIC-EL framework, and one of them has concluded \textit{``Two-way coupled Eulerian–Lagrangian simulations using the
point-force technique have not fulfilled their early promise.''}~\citep{Eaton2009}.

Recent research efforts have focused on improving the accuracy and extending the PSIC-EL method to handle particulate flow simulations which do not satisfy the condition $d_\mathrm{p} \ll h$, by modifying the momentum transfer using convolution with smooth kernels of a certain scale, and by estimating the drag force based on the undisturbed fluid velocity from the disturbed velocity field available on the Eulerian mesh, along with other flow parameters. 
Applying a smooth kernel to the momentum transfer between the particles and the fluid flow can mitigate some of the errors associated with two-way coupling between the fluid and particles~\citep[\textit{e.g.},][]{Capecelatro2013,Evrard2019c,Poustis2019}. Notably, the magnitude of the flow disturbance due to momentum transfer reaches a plateau as the ratio $d_\mathrm{p}/h$ increases, instead of continuing to grow proportionally with this ratio~\citep{Evrard2020a}. The value of this plateau is directly related to the length scale of the regularization kernel, which spreads the transferred momentum over a broader region, resulting in smaller errors.

\begin{figure}
    \centering 
    \begin{tabular}{ccc}
    \subfloat[Disturbed flow, including particle $i$]{\begin{tikzpicture}[x=0.46\textwidth,y=0.46\textwidth]\draw (0,0) node{\adjincludegraphics[width=0.45\textwidth, trim={{.05\width} {.24\height} {.05\width} {.31\height}}, clip, rotate=0]{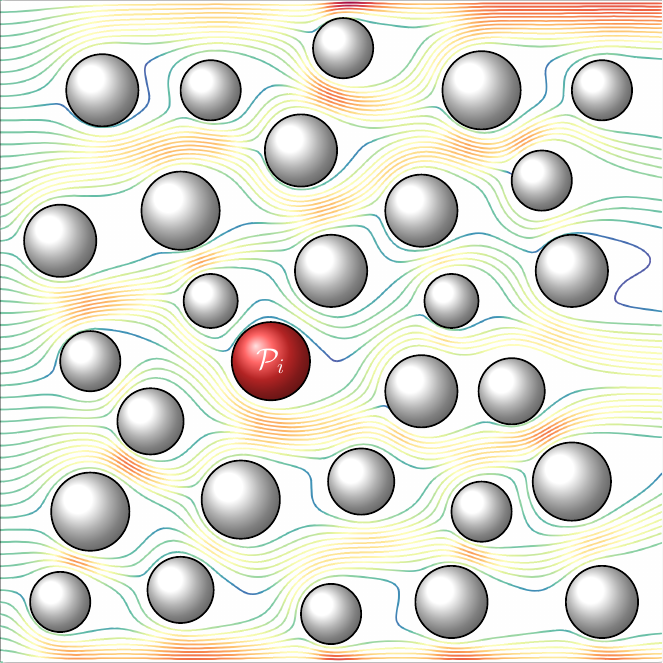} \label{fig:disturbanceflowc}}; 
    \end{tikzpicture}}\vspace{-1mm} & ~ &
    \subfloat[Undisturbed flow, excluding particle $i$]{\begin{tikzpicture}[x=0.46\textwidth,y=0.46\textwidth]\draw (0,0) node{\adjincludegraphics[width=0.45\textwidth, trim={{.05\width} {.24\height} {.05\width} {.31\height}}, clip, rotate=0]{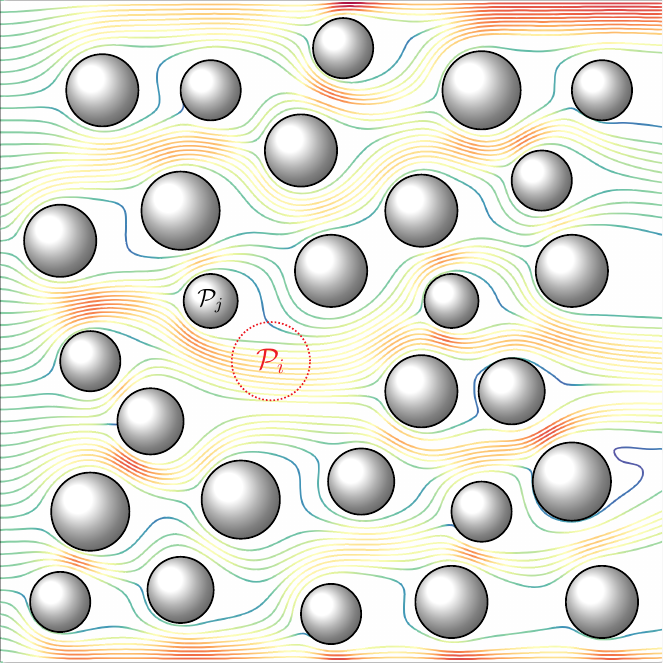} \label{fig:disturbanceflowd}}; 
    \end{tikzpicture}}\vspace{-1mm}
    \end{tabular}
    \caption{The disturbed flow (left) and undisturbed flow (right) for particle $\mathcal{P}_i$ under consideration.}
    \label{fig:disturbanceflow}
\end{figure}

For a given regularization length scale, further error reduction in estimating the drag force necessitates a strategy to improve the estimate of the hydrodynamic forces between the fluid and the particles. 
EL frameworks require an estimate of the relative undisturbed velocity between the fluid and the particle, where the undisturbed fluid velocity is the velocity which the fluid would have at the location of the particle, if the particle under consideration were not present, see Figure~\ref{fig:disturbanceflow}.
To apply this concept, there are a number of recent research works~\citep[\textit{e.g.},][]{Kim2024,Balachandar2022,Evrard2020a,Pakseresht2021,Gualtieri2015,Horwitz2022,Chandran2025}
which focus on recovering the undisturbed fluid velocity from the actual (disturbed) fluid velocity field available, along with other flow parameters. 
Determining the correct hydrodynamic forces in a two-way coupled EL framework by estimating the undisturbed fluid velocity at each particle and subsequently 
using this undisturbed velocity to accurately determine the drag force on the particle using an 
existing drag model has achieved some success, especially in very dilute flows with low particle Reynolds number. However, there 
still exist several fundamental problems to overcome before achieving a general solution following this route.

Firstly, the concept of the undisturbed fluid velocity, as introduced by Maxey, Riley, 
and Gatignol~\citep{Maxey1983,Gatignol1983} may not be the correct quantity to accurately determine 
the hydrodynamic forces. The undisturbed fluid velocity is the fluid velocity in which the 
contribution of the particle under consideration is removed and replaced by the fluid, see Figure~\ref{fig:disturbanceflow}. 
This means that the self-induced flow perturbation by the presence of the particle due to the 
momentum fed back to the fluid by this particle is not present in the undisturbed flow.
For some cases, this undisturbed fluid velocity is easily determined and can be used to 
compute the hydrodynamic forces on the particle. For instance, when a single particle is falling 
in a quiescent fluid, the complete motion of the fluid around the particle is due to the disturbance 
caused by that particle, and the undisturbed fluid velocity should be zero. However, not 
all possible cases are that easy to analyse. For instance, when a particle in a flow crosses its own 
trajectory at a later instance in time, the disturbance from the first instance of the 
presence of the particle should be considered when determining the drag on the particle. 
The situation gets even 
more complicated when multiple particles are present. When estimating the undisturbed 
velocity from a simulation in the absence of the particle, by ``removing'' the particle 
under consideration, the secondary effects of this particle, such as the indirect influences 
occurring on the neighbouring particles, are also removed, which is likely to lead to an 
incorrect drag force prediction. 
This is visualized in Figure \ref{fig:disturbanceflow}, where the fluid velocity field is 
shown with the particle $\mathcal{P}_i$ and without it, \textit{i.e.}, the undisturbed velocity 
field. In the absence of particle $\mathcal{P}_i$, the flow around the neighbouring 
particles is altered, which has an effect on the force experienced by particle $\mathcal{P}_i$.
Typically, models to determine the undisturbed velocity only determine the self-induced velocity disturbance of 
the particle under consideration, which does not consider the secondary effects on 
neighbouring particles. Neglecting these secondary effects is only acceptable in the very dilute regime. In the 
dense regime, force closure models for the estimation of the drag on the particle should 
account for the particle-induced self-disturbance, the effect this disturbance has on the 
neighbouring particles, and finally the neighbour-induced disturbance accounting for 
these secondary effects.

Secondly, accurately determining the undisturbed fluid velocity in practice is generally 
very computationally expensive because the velocity disturbance is generally a result of non-linear 
interactions with the background flow. The general governing equation for the disturbance 
velocity caused by a particle, from which the undisturbed velocity can be obtained, 
is very similar to the Navier-Stokes equations \citep{Evrard2020a}, with a similar 
cost to solve. These equations would need to be solved for every particle in the flow, 
which would be tremendously computationally expensive. Therefore, most approaches to determine the 
undisturbed fluid velocity assume Stokes or Oseen flow 
\citep[\textit{e.g.},][]{Ireland2017,Gualtieri2015, Evrard2020a, Horwitz2022, Chandran2025}, or solve an auxiliary 
set trying to obtain all the particle-induced velocity perturbations as one, in which 
particle-particle effects are neglected \citep[\textit{e.g.},][]{Pakseresht2021}. However, these are all 
approximations of the actual disturbance velocity, and a general solution to this 
problem is likely to be far too computationally expensive in practice.

In the present paper, we propose an alternative framework to accurately obtain the hydrodynamical forces acting on a particle in a flow. This proposed framework does not rely on the undisturbed fluid velocity, but on quantities directly available in an EL framework. A consistent coupling between fluid and particles can be obtained by volume-filtering the Navier-Stokes equations (NSE)~\citep{Hausmann2024a}, a concept introduced by \citet{Anderson1967}. The framework proposed in the present paper links the forces acting on the particles to the local volume-filtered fluid quantities and the ratio of the filter length and the particle diameter, both well-defined in the volume-filtered framework and independent of the particle volume fraction or the fluid mesh resolution. We will first illustrate the novel framework by considering a single particle in a Stokes flow, then extend the framework to a single particle in higher Reynolds number flows, and finally to an assembly of particles in a flow. 

\section{Governing equations}
 We consider the framework of volume-filtering because it allows to derive the concept of EL point-particle simulations directly from the NSE. Volume-filtering a flow quantity $\varPhi$ is defined as the following convolution operation over the whole domain $\Omega$, the union of the fluid domain $\Omega_{\mathrm{f}}$ and the particle domain $\Omega_{\mathrm{p}}$,
\begin{align}
\label{eq:volumefiltering}
    \varepsilon_{\mathrm{f}}(\boldsymbol{x}) \overline{\varPhi}(\boldsymbol{x}) = \int\displaylimits_{\Omega} I_{\mathrm{f}}(\boldsymbol{y}) \varPhi(\boldsymbol{y})g(|\boldsymbol{x}-\boldsymbol{y}|) \mathrm{d}V_y , 
\end{align}
with the radially symmetrical filter kernel, with a length-scale called the filter width, $\sigma$, satisfying
\begin{align}
   \int\displaylimits_{\Omega}g(|\boldsymbol{x}|) \mathrm{d}V_x = 1.
\end{align}
The fluid indicator function, $I_\mathrm{f}$, is defined as
\begin{align}
    I_\mathrm{f}(\boldsymbol{x})=\begin{cases}
        1 & \text{if } \boldsymbol{x} \in \Omega_\mathrm{f}\\
        0 & \text{else }
    \end{cases}
\end{align}
and $\varepsilon_{\mathrm{f}}$ is the fluid volume fraction. The volume-filtered NSE of an incompressible flow with constant density $\rho_{\mathrm{f}}$ and constant dynamic viscosity $\mu_{\mathrm{f}}$ may be written as~\citep{Hausmann2024a}
\begin{align}
\label{eq:continuity}
    \dfrac{\partial \varepsilon_{\mathrm{f}}}{ \partial t} + \dfrac{\partial}{\partial x_i}(\varepsilon_{\mathrm{f}}\Bar{u}_i) &= 0, \\
\label{eq:momentumwithclosures}
    \rho_{\mathrm{f}}\dfrac{\partial \varepsilon_{\mathrm{f}}\Bar{u}_i}{\partial t} + \rho_{\mathrm{f}}\dfrac{\partial}{\partial x_j}(\varepsilon_{\mathrm{f}}\Bar{u}_i\varepsilon_{\mathrm{f}}\Bar{u}_j) &= -\dfrac{\partial \varepsilon_{\mathrm{f}}\Bar{p}}{\partial x_i}+\mu_{\mathrm{f}} \dfrac{\partial^2\varepsilon_{\mathrm{f}}\Bar{u}_i}{\partial x_j \partial x_j} - \sum_q s_{q,i}+ \mu_{\mathrm{f}} \mathcal{E}_i -\rho_{\mathrm{f}}\dfrac{\partial }{\partial x_j}\tau_{\mathrm{sfs},ij},
\end{align}
where $u_i$ is the fluid velocity, $p$ is the fluid pressure,
$\mathcal{E}_i$ represents the viscous closure, and $\tau_{\mathrm{sfs},ij}$ the subfilter stress tensor. $\mathcal{E}_i$ can be expressed analytically, and $\tau_{\mathrm{sfs},ij}$ requires modelling and is important to take into account for larger Reynolds numbers~\citep{Hausmann2024a}. The particle momentum source is defined as the following sum of integrals over the surfaces of the particles with index $q$
\begin{align}
    s_i = \sum_q \int\displaylimits_{\partial\Omega_{\mathrm{p},q}}g(|\boldsymbol{x}-\boldsymbol{y}|)\left(-p \delta_{ij} + \mu_{\mathrm{f}} \left(\dfrac{\partial u_i}{\partial y_j}+\dfrac{\partial u_j}{\partial y_i}\right)\right)n_j\mathrm{d}A_y,
\end{align}
where $n_j$ is the normal vector at the surface of the particle. Since $s_i$ depends on the unfiltered fluid velocity and pressure fields, it typically requires modelling. For large filter widths, $\sigma$, to particle diameter, $d_\mathrm{p}$, ratios, as it is common for point-particle simulations, the particle momentum source is typically approximated as (see, \textit{e.g.}, \citet{Capecelatro2013})
\begin{align}
    s_i \approx \sum_q F_{\mathrm{h},q,i} \, g(|\boldsymbol{x} - \boldsymbol{x}_{\mathrm{p},q}|),
\end{align}
which has been shown in \citet{Hausmann2024a} to be a reasonable assumption for $\sigma/d_{\mathrm{p}}\ge 1$. Here, $F_{\mathrm{h},i}$ is the hydrodynamic force on the particle. In this work, we assume that the dominant hydrodynamic force on the particle is the drag force, $\mathbf{F}_\mathrm{h} \approx \mathbf{F}_\mathrm{d}$, although the framework can be extended to account for other hydrodynamic forces as well.
The drag model is a crucial aspect when predicting the motion of particles in a fluid flow. It encompasses a wide range of complexities, from the basic linear drag law for an isolated particle in Stokes flow, to more sophisticated formulations that consider various flow regimes and particle interactions.

\section{A novel class of force correlations based on the volume-filtered fluid velocity}

In the drag force models used in EL frameworks, the free-stream fluid velocity, 
$\boldsymbol{u}_{\infty}$, is typically interpreted as the undisturbed fluid velocity. Additionally, in some cases,
self-induced velocity disturbance correction models are used to determine $\boldsymbol{u}_{\infty}$ from 
the volume-filtered fluid velocity which is available in the Euler-Lagrange point-particle 
simulation. The drawbacks of this procedure have been discussed in section \ref{sec:introduction}. 
We propose a novel class of force correlations, in which the hydrodynamic force depends directly on the volume-filtered fluid velocity at the particle position and the filter width
as inputs. This requires changing existing force correlations, or even deriving new force correlations. 

The principle will be first shown for a single particle in Stokes flow and a single particle in a finite Reynolds number flow, and will also be derived for a flow past particles in assemblies with varying volume fractions.

\subsection{A sphere in Stokes flow}
Stokes drag on a sphere is the simplest form of drag force determination, applicable to small spherical particles moving at low particle Reynolds numbers, $\mathrm{Re_p} \ll 1 $, where inertial effects are negligible compared to viscous forces.  The particle Reynolds number is the Reynolds number based on the particle diameter and the relative velocity between the fluid and the particle. The expression for Stokes drag for a uniform flow in an infinite domain is analytically given as
\begin{equation}
\label{eq:stokesdrag}
\boldsymbol{F}_\mathrm{d} = 3 \pi \mu_{\mathrm{f}} d_\mathrm{p} \boldsymbol{U}_\mathrm{rel},
\end{equation}
where $\boldsymbol{U}_\mathrm{rel} =(\boldsymbol{u}_{\infty}-\boldsymbol{v})$ is the relative velocity between the particle and the uniform fluid velocity very far away from the particle, $\boldsymbol{u}_{\infty}$, herein referred to as the undisturbed velocity. The particle velocity is written as $\boldsymbol{v}$.
This linear relationship between the drag force and undisturbed velocity indicates that the drag force is directly proportional to the particle velocity. Stokes drag is derived under the assumption of steady, laminar flow with no significant effects from the particle wake.

In order to obtain the drag force dependence on the volume-filtered velocity at the particle 
centre, $\varepsilon_{\mathrm{f}}\Bar{\boldsymbol{u}}_{f@p}$, and the non-dimensional relative filter width, $\sigma^\prime = \sigma/d_\mathrm{p}$, the analytical velocity, $\boldsymbol{u}$,
of the Stokes flow past a sphere moving with a velocity $\boldsymbol{v}$ is volume-filtered
. This can be written as follows:

\begin{equation}
        \varepsilon_{\mathrm{f}}\Bar{\boldsymbol{u}}_{f@p} = 2 \pi \int_0^{\pi} \int_{d_{\mathrm{p}}/2}^{\infty} (\boldsymbol{u} + \boldsymbol{v}) ~g ~r^2 \sin\theta ~\mathrm{d}r ~\mathrm{d}\theta,
    \label{eq:integral_gaussianfiltering}
\end{equation}
where $g$ is the Gaussian distribution function with standard deviation $\sigma$. It should be noted that $\boldsymbol{u}$ is defined in the reference frame of the moving particle in spherical coordinates \citep{Batchelor1967}.  Utilizing the symmetry of $\boldsymbol{u}$ in the azimuthal direction, and integrating $\boldsymbol{u}$ along the polar ($\theta$) and radial ($r$) directions, we obtain:
\begin{equation}
        \varepsilon_{\mathrm{f}}\Bar{\boldsymbol{u}}_{f@p} = (\boldsymbol{u}_{\infty} - \boldsymbol{v}) \erfc{ \left(
        \frac{1}{2 \sqrt{2} \sigma^\prime} \right) } + \varepsilon_{\mathrm{f}} \boldsymbol{v}.
    \label{eq:gaussianfilteredvelocity}
\end{equation}

Substituting ($\boldsymbol{u}_{\infty} - \boldsymbol{v}$) from the Equation above into Equation~\eqref{eq:stokesdrag}, 
the drag force on the particle as a function of the local filtered velocity is as follows:
\begin{equation}
\label{eq:filteredstokesdrag}
\boldsymbol{\tilde{F}}_\mathrm{d} = \dfrac{3 \pi \mu d_\mathrm{p} (\varepsilon_{\mathrm{f}}\Bar{\boldsymbol{u}}_{f@p}-\varepsilon_{\mathrm{f}}\boldsymbol{v})}{\erfc{\left(
 \frac{1}{2 \sqrt{2} \sigma^\prime} \right) }
}.
\end{equation}
This result is similar to the drag law proposed by \cite{Ireland2017}, although the steps to derive it are significantly different.

In the limit of $\sigma^\prime \to \infty$, 
$\varepsilon_{\mathrm{f}}\Bar{\boldsymbol{u}}_{f@p} \to \boldsymbol{u}_{\infty}$ and 
Equation \eqref{eq:filteredstokesdrag} converges to the standard Stokes drag 
defined in Equation~\eqref{eq:stokesdrag}. The drag force on the particle, 
$\boldsymbol{\tilde{F}}_\mathrm{d}$ is a function of the volume-filtered 
velocity at the particle centre, a quantity that can be directly interpolated 
from the computational grid in a volume-filtered EL simulation, and the relative filter width.

\subsection{A sphere in finite Reynolds number flow}
As the particle Reynolds number increases beyond the Stokes regime, the drag force no longer remains proportional to the relative velocity. To determine the drag force for particles with intermediate particle Reynolds numbers, 1 $<$ $\mathrm{Re}_\mathrm{p}$ $<$ 1000, empirical correlations are often used to account for the increased non-linearity in the drag force. One common empirical formula is the correlation of \citet{Schiller1933}, which proposes a drag coefficient $C_\mathrm{D}$ of
\begin{equation}
\label{eq:CDSchillerNaumann}
C_\mathrm{D} = \frac{24}{\mathrm{Re_p}} \left(1 + 0.15 \, \mathrm{Re_p}^{0.687}\right),
\end{equation}
and the drag force is then given by
\begin{equation}
\label{eq:dragforce}
\boldsymbol{F}_\mathrm{d} = \frac{1}{2} C_\mathrm{D} 
 \, \rho_\mathrm{f} \, d^2_\mathrm{p} \left| \boldsymbol{U}_\mathrm{rel} \right| \boldsymbol{U}_\mathrm{rel},
\end{equation}
where $\rho_\mathrm{f}$ is the density of the fluid. 

Similar to the model proposed for a sphere in Stokes flow, we now seek a relation between $\boldsymbol{u}_{\infty}$ and the volume-filtered velocity at the center of the particle, $\varepsilon_{\mathrm{f}}\Bar{\boldsymbol{u}}_{f@p}$, for a uniform flow around a sphere at particle Reynolds numbers larger than zero. In this case, it is impossible to derive an analytical relation between $\boldsymbol{u}_{\infty}$ and $\varepsilon_{\mathrm{f}}\Bar{\boldsymbol{u}}_{f@p}$, as the fluid velocity field around a particle is not known analytically. 
In order to derive an empirical relation between 
$\boldsymbol{u}_{\infty}$ and $\varepsilon_{\mathrm{f}}\Bar{\boldsymbol{u}}_{f@p}$ for the uniform flow around a sphere at finite Reynolds numbers, we carry out particle-resolved direct numerical simulations (PR-DNS) of a sphere at Reynolds numbers in the range $1\le\mathrm{Re_p}\le200$.  

\begin{figure}
    \centering
    \includegraphics[width=0.9\linewidth]{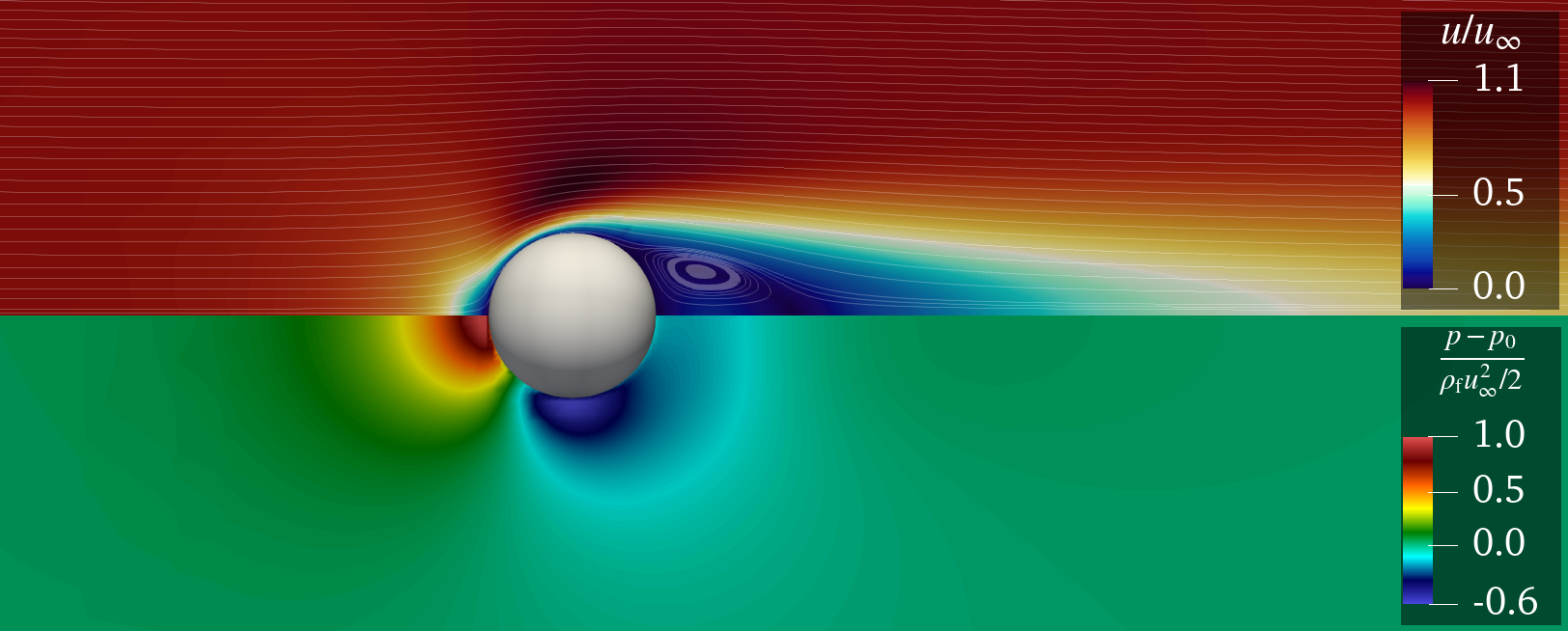}
    \caption{Flow past a single particle at $\mathrm{Re_p}=100$. Top half is coloured by the flow velocity normalized by the free-stream velocity, $u_\infty$, along with streamlines. Bottom half is coloured by the pressure normalized by $\rho_\mathrm{f} u_\infty^2 / 2$.}
    \label{fig:single_particle}
\end{figure}

Our PR-DNS framework employs a finite-volume approach to solve the incompressible Navier-Stokes equations with second-order accurate spatio-temporal discretization over a mesh. The flow is driven by a body force in the direction of the primary flow, while no-slip and no-penetration boundary conditions at the particle surface are enforced using a momentum source term computed via the hybrid immersed boundary method~\citep{Cheron2023a}. The fluid governing equations are solved numerically on a grid which is refined near the surface of the particle, achieving a resolution of $d_\mathrm{p}/\Delta x \approx 36$, where $\Delta x$ is the cell size of the grid near the particle surface. In Figure \ref{fig:single_particle}, a visualization of the flow past an isolated particle is shown for the case of particle Reynolds number $\mathrm{Re_p}=100$. 

The volume-filtered fluid velocity at the center of the particle is obtained by explicitly volume-filtering the velocity field obtained from the PR-DNS, which is done in two steps. Firstly, the velocity field is multiplied with the fluid indicator function, \textit{i.e.}, fluid velocities in the mesh cells inside the particle are multiplied with zero. Secondly, the convolution integral for the volume-filtered velocity at the center of the particle is computed discretely as
\begin{align}
    \varepsilon_{\mathrm{f}}(\boldsymbol{x}_\mathrm{c}) \overline{\boldsymbol{u}}(\boldsymbol{x}_\mathrm{c}) = \int\displaylimits_{\Omega} I_{\mathrm{f}}(\boldsymbol{y}) \boldsymbol{u}(\boldsymbol{y})g(|\boldsymbol{x}_\mathrm{c}-\boldsymbol{y}|) \mathrm{d}V_y \approx \sum_{n=1}^{N_{\mathrm{f}}} I_{\mathrm{f},n} \boldsymbol{u}_n\tilde{g}_n\Delta V_n,
\end{align}
where $ I_{\mathrm{f},n}$ and $\boldsymbol{u}_n$ are the values of the fluid indicator function and the velocity in mesh cell $n$, $\Delta V_n$ is the volume of the mesh cell, $\tilde{g}_n$ is the integral contribution of the Gaussian to the mesh cell, and $N_{\mathrm{f}}$ is the total number of mesh cells in a simulation. 

From explicitly volume-filtering the PR-DNS results, pairs of $\boldsymbol{u}_{\infty}$ and $\varepsilon_{\mathrm{f}}\Bar{\boldsymbol{u}}_{f@p}$ for Reynolds numbers in the range of $\mathrm{{Re}_p}\in [1,10,50,100,200]$ and for the relative filter widths $\sigma^\prime \in [0.5,1,2,3,4,5]$ are obtained. Note that for values of relative filter widths of $\sigma^\prime >5$,  $\varepsilon_{\mathrm{f}}\Bar{\boldsymbol{u}}_{f@p}$ no longer changes significantly compared to its value obtained with a relative filter width of $5$.

The relation between $\boldsymbol{u}_{\infty}$ and $\varepsilon_{\mathrm{f}}\Bar{\boldsymbol{u}}_{f@p}$ must satisfy two asymptotic limits: (\romannumeral 1) Equation~ \eqref{eq:gaussianfilteredvelocity} must be recovered as $\mathrm{\overline{Re}_p}\rightarrow 0$, and (\romannumeral 2) $\varepsilon_{\mathrm{f}}\Bar{\boldsymbol{u}}_{f@p} \rightarrow \boldsymbol{u}_{\infty}$ as $\sigma^\prime \rightarrow \infty$. We propose the following empirical correlation that satisfies these asymptotic limits:
\begin{align}
    \label{eq:correlationforurel}
    \boldsymbol{U}_{\mathrm{rel}} = \boldsymbol{u}_\infty - \boldsymbol{v} = \dfrac{\varepsilon_{\mathrm{f}}\Bar{\boldsymbol{u}}_{f@p} - \varepsilon_{\mathrm{f}}\boldsymbol{v}}{\erfc{ \left(
 \frac{1}{2 \sqrt{2} \sigma^\prime} \right)}
    }(1+k_\sigma k_\mathrm{Re}),
\end{align}
with 
\begin{align}
    k_\sigma = \frac{1}{2}\left(\frac{a_0 (\sigma^{\prime}-0.5)^{a_1}}{1+a_0(\sigma^{\prime}-0.5)^{a_1}} -1 \right),
\end{align}
and 
\begin{align}
    k_\mathrm{Re} = \frac{1}{2}\left(1 + \mathrm{erf}(a_2 \log_{10}(\mathrm{\overline{Re}_{\mathrm{p}}}) - a_3) \right),
\end{align}
where $\mathrm{\overline{Re}_{\mathrm{p}}}=\rho_\mathrm{f} |\varepsilon_{\mathrm{f}}\Bar{\boldsymbol{u}}_{f@p} - \varepsilon_{\mathrm{f}}\boldsymbol{v}|d_\mathrm{p}/\mu_\mathrm{f}$ is the filtered particle Reynolds number. The coefficients of the empirical correlation are given in table \ref{tab:parameter}. After obtaining $\boldsymbol{U}_{\mathrm{rel}}$, the drag force on the particle can then be computed with Equation~\eqref{eq:dragforce}. 

\begin{figure}    \centering
    \includegraphics[scale=0.86]{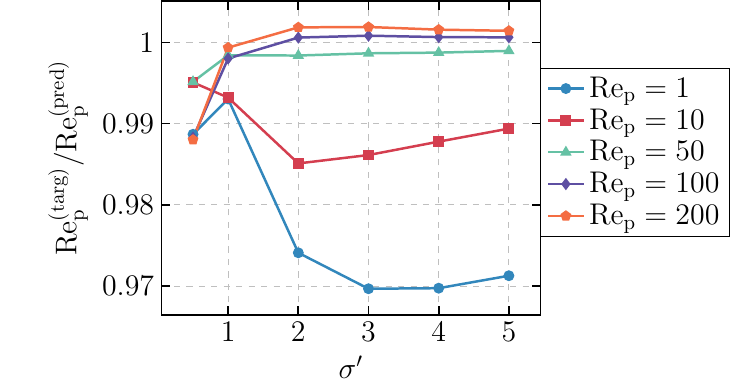}
    \caption{Target undisturbed Reynolds number, $\mathrm{Re}_\mathrm{p}^\mathrm{(targ)}$, divided by the predicted undisturbed Reynolds number, $\mathrm{Re}_\mathrm{p}^\mathrm{(pred)}$, with the empirical correlation for $\boldsymbol{U}_{\mathrm{rel}}$ as given in Equation~ \eqref{eq:correlationforurel} for different normalized filter widths $\sigma^\prime$.}
    \label{fig:errorschillernaumann}
\end{figure}

\begin{table}
  \begin{center}
\def~{\hphantom{0}}
  \begin{tabular}{p{1.0cm}p{1.0cm}p{1.0cm}p{1.0cm}}
      $a_0$ & $a_1$ & $a_2$ & $a_3$ \\[3pt]
      1.1076&1.0359& 0.8220& 0.2135
  \end{tabular}
  \caption{Coefficients for the empirical correlation to determine $\boldsymbol{U}_{\mathrm{rel}}$ for finite $\mathrm{Re_p}$, which is given in Equation~\eqref{eq:correlationforurel}.}
  \label{tab:parameter}
  \end{center}
\end{table}
Figure \ref{fig:errorschillernaumann} shows the target undisturbed Reynolds number, $\mathrm{Re}_\mathrm{p}^\mathrm{(targ)}$, divided by the predicted undisturbed Reynolds number, $\mathrm{Re}_\mathrm{p}^\mathrm{(pred)}$, computed with the empirical correlation for $\boldsymbol{U}_{\mathrm{rel}}$ for different Reynolds numbers. The maximum deviation of approximately 3\% occurs at a Reynolds number of 1 for a relative filter width of 3. We assume this deviation to be significantly smaller than the modeling and discretization errors that typically arise in EL point-particle simulations.

\subsection{Suspension of monodisperse spheres}
\label{ssec:correlationsuspension}
In particle-laden flows of practical relevance, the individual particles do not experience a uniform flow field in an unbounded domain; instead, the presence of surrounding particles alters the local flow so it becomes non-uniform, thereby also influencing the hydrodynamic forces acting on each particle.  There are a number of proposed  empirical models \cite[\textit{e.g.},][]{Wen1966,Gidaspow1986,Beetstra2007,Tenneti2011} to predict the drag force on a particle in a suspension. 
These models are typically based on the spatial mean of the relative velocity between the fluid and the particle, the mean particle volume fraction, and the properties of the particle. 
For instance, in \cite{Tenneti2011}, PR-DNS of various particle assemblies in periodic domains are carried out, and the corresponding forces on the particles in the assemblies are averaged and are used to propose an empirical expression for the drag force, which depends on the ``global'' fluid velocity and the ``global'' volume fraction only, where ``global'' refers to the average over the periodic domain size in which the simulations are carried out.

However, because the flow around each particle surrounded by other particles can be complex and can vary significantly between neighbouring particles, the drag force between the particles can also strongly vary, and the aforementioned force correlations cannot produce an accurate value for each particle.
More recently, models have been proposed that aim to predict the deviation of the hydrodynamic force acting on an individual particle \cite[\textit{e.g.},][]{Akiki2017a,Hardy2022,vanWachem2024}. The drag force predicted by these models also depends
on the location of the neighbouring particles, in some way. This approach is suitable for flows with a homogeneous particle distribution and relatively large filter widths. However, the definitions of the superficial velocity and global particle volume fraction become ambiguous in inhomogeneous flows, such as flows with particle clustering, since the value of both quantities depends on how large the averaging volume is, which is an arbitrary choice. 

The presently proposed framework, in which the force correlations depend on volume-filtered quantities, can be extended to particle suspensions. Since the flow is more complicated than the uniform flow around a single particle, additional uniquely defined volume-filtered quantities that are accessible in EL simulations have to be considered to predict the hydrodynamic force accurately. 
Independent of the complexity of the particle-laden flow, the hydrodynamic force on particle $q$ is 
expressed as a function of the pressure and velocity field as
\begin{align}
    F_{\mathrm{h},q,i} = \int\displaylimits_{\partial\Omega_{\mathrm{p},q}} \left[-p \delta_{ij} + \mu_{\mathrm{f}} \left(\dfrac{\partial u_i}{\partial x_j}+\dfrac{\partial u_j}{\partial x_i}\right)\right] n_j\mathrm{d}A_x,
\end{align}
if the values of $u_i$ and $p$ are known on all points on the surface of the particle.
Formally, the process of volume-filtering does not remove any information, but the discretization of the filtered solution on a finite fluid mesh does. Therefore, an exact relation between the volume-filtered flow quantities and the hydrodynamic force exists, but since the volume-filtered flow quantities are approximated in EL simulations, the predicted hydrodynamic force is also an approximation. Since less information is removed for small filter widths, it is expected that the estimation of the hydrodynamic force is also more accurate for small filter widths. 

As a conceptual proof of the proposed framework for force correlations, the functional approach of the existing mean drag force correlation of \citet{Tenneti2011} is adapted to the framework proposed in this paper. The adaption consists of two essential modifications: (\romannumeral 1) The input parameters are the volume-filtered velocity and the volume fraction at the particle position as well as the filter width, instead of the superficial velocity and the global volume fraction, as in the original model. (\romannumeral 2) The coefficients of the expression are obtained by minimizing the deviation of the predicted force from the actual force of each individual particle, instead of the deviation of the predicted force from the mean of the forces acting on all the considered particles in each PR-DNS of a particle assembly. 

To determine the coefficients of the proposed expression, we have carried out PR-DNS of assemblies of  particles. The PR-DNS are performed in a periodic cubic domain of volume $V_\mathrm{tot}$, containing $N_\mathrm{p}$ non-overlapping, monodisperse spherical particles. The domain is discretized into $N_\mathrm{f}$ Eulerian fluid cells, in which the governing equations for fluid flow are solved to obtain the local fluid properties. Each fluid cell may be fully occupied by fluid, fully occupied by a particle, or partially occupied by both. Flow through the periodic domain is driven by imposing a pressure drop, $\delta P$, which is introduced as an additional body force in the $x$-direction in the fluid momentum equations, so the correct superficial fluid velocity is obtained. For further details, we refer to \cite{vanWachem2024}.

In this study, a total of six global particle volume fractions are considered, namely $\langle \varepsilon_\mathrm{p}\rangle$ = $0.1$, $0.2$, $0.3$, $0.4$, $0.5$, and $0.6$. For each global particle volume fraction, multiple flow conditions are investigated by varying the superficial particle Reynolds number as $\mathrm{Re}_\mathrm{s} = 0.1$, $1$, $10$, $50$, $100$, and $300$. The superficial particle Reynolds number is defined as $\mathrm{Re}_\mathrm{s}=\rho_\mathrm{f}d_\mathrm{p} u_\mathrm{s}/\mu_\mathrm{f}$, where the superficial velocity is given as the fluid domain average of the fluid velocity,
\begin{align}
    u_\mathrm{s} = \dfrac{1}{V_\mathrm{tot}} \int\displaylimits_{\Omega_\mathrm{f}} u_x(\mathbf{x})\mathrm{d}V_x.
\end{align}
The total volume of the domain, $V_\mathrm{tot}$, is the sum of the volume occupied by the fluid and the volume occupied by the particles. To ensure statistical convergence, three independent realizations are simulated for each combination of solid volume fraction and superficial particle Reynolds number. Each simulation is carried out until a statistically steady state is achieved. In total, 108 PR-DNS are performed, with an average of 136.33 particles per simulation.
This results in a total of 14,724 data points across the entire parameter space, which serve as the basis for the development of the hydrodynamic force correlations.

From the results of the PR-DNS, we determine the volume-filtered flow quantities for various filter widths, for each simulation configuration.
To obtain the volume-filtered flow quantities, we exploit the fact that the domain is periodic and that a convolution becomes a multiplication in spectral space. Therefore, the volume-filtered velocity is given as
\begin{align}
    \varepsilon_\mathrm{f}\overline{\boldsymbol{u}} = g\ast (I_\mathrm{f}\boldsymbol{u}) = \mathcal{F}^{-1}\{ \mathcal{F}\{ g\}\mathcal{F}\{I_\mathrm{f}\boldsymbol{u}\} \},
\end{align}
where $\ast$ is the convolution operator, and $\mathcal{F}$ corresponds to the Fourier transform. 

In order to derive a drag correlation as a function of the local volume-filtered quantities, we propose a drag correlation of a similar form as the one in \citet{Tenneti2011}, except that the input parameters here are the local fluid volume fraction, the volume-filtered velocity, and the ratio of the filter width to the particle diameter, $\sigma^{\prime}$. The general form of the normalized drag correlation is as follows:
\begin{align}
    \boldsymbol{\tilde{F}}_\mathrm{d} &=
    3 \pi \mu d_{\mathrm{p}} (\varepsilon_{\mathrm{f}} \Bar{\boldsymbol{u}}_\mathrm{f@p}  - \varepsilon_{\mathrm{f}}\boldsymbol{v})
    \left[ \frac{C_{\mathrm{D}} \, \mathrm{\overline{Re}_{\mathrm{p}}}}{24} \frac{1}{(1 - \delta_{\varepsilon})^3} + a_0 \frac{\delta_{\varepsilon}}{(1 - \delta_{\varepsilon})^3} \right. \nonumber \\  
    &  \left. + a_1 \frac{\delta_{\varepsilon}^{1/3}}{(1 - \delta_{\varepsilon})^4} + \delta_{\varepsilon}^{a_4} \mathrm{\overline{Re}_{\mathrm{p}}} \left(a_2 + a_3 \frac{\delta_{\varepsilon}^{a_5}}{(1-\delta_{\varepsilon})^2} \right) \right].
    \label{eq:filtered_dragcorrelation}
\end{align}
$\delta_{\varepsilon}$ is the difference between the fluid volume fraction of an isolated particle and $\varepsilon_\mathrm{f}$, both of which are evaluated at the particle centre. For a Gaussian kernel, $\delta_{\varepsilon}$
 can be mathematically expressed as,
\begin{equation}
    \delta_{\varepsilon} = \erfc \left(\frac{1}{2 \sqrt{2} \sigma^\prime} \right) + \frac{\exp\left(-\frac{1}{8 {\sigma^\prime}^2}\right)}{\sigma^\prime \sqrt{2 \pi}} - \varepsilon_{\mathrm{f}}.
    \label{eq:delta_localvfrac_defn}
\end{equation}
The equation above is obtained by using the fluid volume fraction for an isolated particle \citep{Balachandar2022}, and using L'H\^opital's rule to retrieve the volume fraction at the particle centre. In the limit of a very dilute volume fraction, \textit{i.e.}, $\delta_{\varepsilon} \to 0$, Equation \eqref{eq:filtered_dragcorrelation} reduces to the standard Schiller-Naumann correlation for a single particle. Mean drag force models, such as the one proposed by \citet{Tenneti2011}, use the ``global'' superficial fluid velocity, and the global particle volume fraction to predict a mean drag force in an assembly. When $\sigma^{\prime} \to \infty$, $\delta_{\varepsilon}$ simplifies to the global particle volume fraction, $\langle \varepsilon_{\mathrm{p}} \rangle$, and the volume-filtered velocity at the particle centre simplifies to the superficial fluid velocity, $u_\mathrm{s}$. In this limit, Equation \eqref{eq:filtered_dragcorrelation} uses the same inputs as in a mean drag force model, such as the one proposed by \citet{Tenneti2011}.
 
To evaluate the coefficients in Equation \eqref{eq:filtered_dragcorrelation}, we define an error measure that minimizes the deviation of the drag force prediction from the actual drag experienced by each individual particle in the random assembly. The mean relative deviation of the predicted drag force in the streamwise direction acting on particle $q$, $\tilde{F}_{\mathrm{d},q,x}$, from the actual drag force acting on particle $q$, $F_{\mathrm{d},q,x}$, is therefore defined as
\begin{align}
    {E}_\mathrm{F} = \dfrac{1}{N_\mathrm{p}} 
    \sum_{q=1}^{N_\mathrm{p}} \bigg| \dfrac{F_{\mathrm{d},q,x} - \tilde{F}_{\mathrm{d},q,x}}{F_{\mathrm{d},q,x}} \bigg| .
    \label{eq:errormeasure_filtereddragcorrelation}
\end{align}

Table \ref{tab:parameter_filtered_tenneti} presents the fitted coefficients of Equation \eqref{eq:filtered_dragcorrelation} obtained by minimizing the mean deviation in the predicted individual particle force against the corresponding particle force from the PR-DNS data using Equation~\eqref{eq:errormeasure_filtereddragcorrelation}. The datasets are grouped based on $\sigma^\prime$, and use the L-BFGS-B algorithm for limited memory bounded-constraint optimization to evaluate the coefficients \citep{Byrd1996}. A maximum mean error of approximately 27~\% in predicting the individual particle forces across the six different values of $\sigma^\prime$ is observed.

\begin{table}
  \begin{center}
\def~{\hphantom{0}}
  \begin{tabular}{p{1.0cm}p{1.0cm}p{1.0cm}p{1.0cm}p{1.0cm}p{1.0cm}p{1.0cm}}
      $\sigma / {d_{\mathrm{p}}}$ & $a_0$ & $a_1$ & $a_2$ & $a_3$ & $a_4$ & $a_5$ \\ [3pt]
      0.5 & 9.147 & 9.955 & 0.077 & 8.266 & 0.056 & 2.898 \\ [3pt]
      1   & 4.589 & 2.122 & 0.097 & 2.308 & 0.604 & 3.571 \\ [3pt]
      2   & 6.486 & 0.609 & 0.109 & 1.598 & 0.891 & 3.274 \\ [3pt]
      3   & 7.501 & 0.190 & 0.131 & 1.699 & 1.012 & 3.451 \\ [3pt]
      4   & 7.584 & 0.134 & 0.550 & 5.393 & 1.826 & 5.097 \\ [3pt]
      5   & 7.615 & 0.101 & 0.724 & 8.188 & 2.003 & 5.856 \\ [3pt]
  \end{tabular}
  \caption{Coefficients of the empirical correlation for $\boldsymbol{\tilde{F}}_\mathrm{d}$ given in Equation \eqref{eq:filtered_dragcorrelation}.}
  \label{tab:parameter_filtered_tenneti}
  \end{center}
\end{table}

\section{Results and discussion}
\subsection{A sphere falling in a fluid in the Stokes regime}
We demonstrate the advantages of the newly proposed framework of force correlations by first considering a single falling spherical particle in a large domain filled with an initially quiescent  fluid under gravity in the Stokes regime. The particle, which is initially at rest, accelerates in the fluid until it reaches its terminal velocity, which can be determined analytically in this regime. We perform three configurations of two-way coupled EL point-particle simulations of this case by:
\begin{enumerate}
    \item Solving the volume-filtered NSE with the classical Stokes drag,
    \item Solving the volume-filtered NSE with the newly proposed filtered Stokes drag, equation \eqref{eq:filteredstokesdrag}, and
    \item Solving the commonly used PSIC method~\citep{Crowe1977} and neglecting
    $\mathcal{E}_i$ and $\tau_{\mathrm{sfs},ij}$.
\end{enumerate}

For each configuration, the domain is fully periodic and has a size of $L_x\times L_y\times L_z = 100d_{\mathrm{p}}\times 100d_{\mathrm{p}}\times 100d_{\mathrm{p}}$ and various mesh resolutions and, where applicable, filter widths are simulated. The particle Reynolds number based on the terminal velocity is $\mathrm{Re_p}=1.11\times10^{-4}$ and the density ratio is $\rho_\mathrm{p}/\rho_\mathrm{f}=2000$.

\begin{figure}
    \centering
    \includegraphics[scale=0.56]{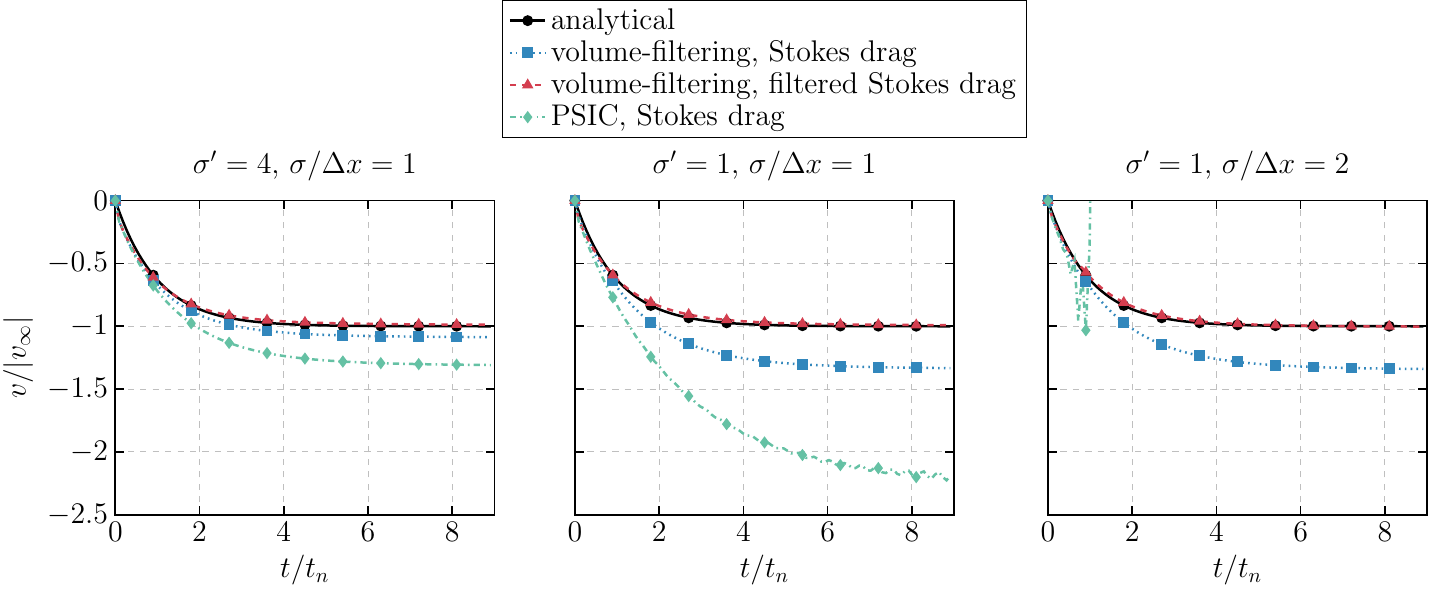}
    \caption{Relative particle velocities of a single falling sphere over time for the volume-filtered simulations using classical Stokes drag, the volume-filtered simulations with the filtered Stokes drag, and the particle-source-in-cell (PSIC) simulation frameworks, compared to the analytical solution. The simulations are performed with three different resolutions, $d_{\mathrm{p}}/\Delta x = 0.25$, $d_{\mathrm{p}}/\Delta x = 1$, and $d_{\mathrm{p}}/\Delta x = 2$ (from left to right). }
    \label{fig:sedimentation}
\end{figure}

Figure \ref{fig:sedimentation} shows the particle velocities of the three different simulation configurations for three different resolutions, $d_{\mathrm{p}}/\Delta x = 0.25$, $d_{\mathrm{p}}/\Delta x = 1$, and $d_{\mathrm{p}}/\Delta x = 2$, and in the figure the simulation results are compared with the analytical result. The filter widths, $\sigma$, are chosen according to the guidelines provided in \citet{Hausmann2024a}, such that $\sigma^\prime \ge 1$ and $\sigma/\Delta x \ge 1$ are always satisfied. 

The simulations using the classical Stokes drag overpredict the magnitude of the terminal velocity considerably, although the simulations do reach a stable velocity after some time. The simulations adopting the PSIC framework show an even larger error, and do not even converge for $d_{\mathrm{p}}/\Delta x = 2$. When using the newly proposed drag force correlation based on the volume-filtered velocity and the ratio of the filter width and the particle diameter, the analytical particle velocity is reproduced accurately in all cases.

\subsubsection{Comparison with undisturbed velocity models}
A falling sphere in a finite domain is often used to validate models based on the undisturbed fluid velocity principle.
The most recent velocity disturbance corrections for transient flow \citep{Evrard2025,Chandran2025} can predict the trajectory of a single falling sphere in a finite domain accurately because this is one of the few cases in which the undisturbed velocity is well-defined. We investigate a single falling particle under gravity in a small periodic domain, a simple case where the concept of subtracting the velocity disturbance to achieve the undisturbed fluid velocity fails when the particle is
confronted with its historic effect on the fluid.

\begin{figure}
    \centering
    \includegraphics[scale=0.6]{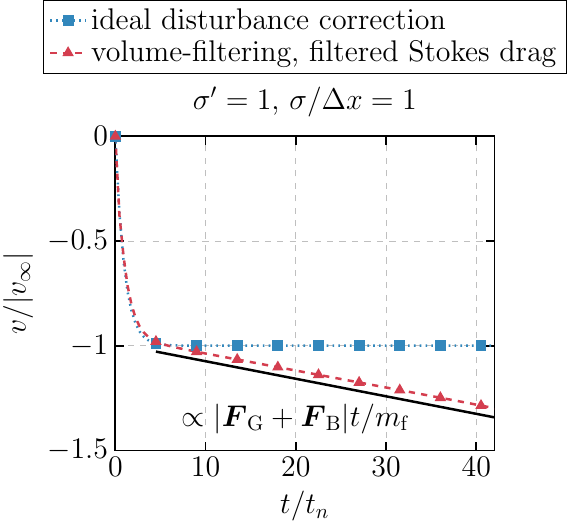}
    \caption{A falling isolated particle in a periodic domain. Simulations of the volume-filtered NSE using the proposed filtered Stokes drag are compared to the theoretical acceleration of the particle and the ideal correction of the flow disturbance of the particle.}
    \label{fig:periodicsedimentation}
\end{figure}

To simulate this case, two-way coupling is used, \textit{i.e.}, the drag force acting on the particle is fed back to the fluid momentum. This results in a constant acceleration of the mean fluid velocity. Since in steady state the drag force on the particle must be equal to the sum of the  gravitational force, $\boldsymbol{F}_{\mathrm{G}}$, and the buoyancy force, $\boldsymbol{F}_{\mathrm{B}}$, the mean flow accelerates with a magnitude of $|\boldsymbol{F}_{\mathrm{G}}+\boldsymbol{F}_{\mathrm{B}}|/m_{\mathrm{f}}$, where the mass of the fluid, $m_{\mathrm{f}} = \rho_{\mathrm{f}}L_xL_yL_z$. The periodic domain has a size of $L_x\times L_y\times L_z = 50d_{\mathrm{p}}\times 50d_{\mathrm{p}}\times 50d_{\mathrm{p}}$. A steady state is achieved only in the sense of the relative velocity between the particle and the fluid, as the mean fluid flow continues to accelerate, just as the particle. However, the particle velocity in the non-moving Eulerian frame of reference continues to increase in time. The results of the EL simulation using the concept of the undisturbed fluid velocity with the EL simulation using the volume-filtered drag model, as well as the theoretically expected trend, is shown in Figure~\ref{fig:periodicsedimentation}. It is clearly evident that, in this case, the concept of reconstructing the undisturbed velocity to compute the drag force fails to produce accurate results, whereas the simulation employing the volume-filtered drag force framework yields predictions in good agreement with the reference solution.

One could argue that a single particle which is falling in a periodic domain does not have much practical relevance. However, the failure of the concept of velocity disturbance correction directly translates to configurations with more than one particle, which is frequently studied in the literature (see, \textit{e.g.}, \citet{Uhlmann2014,Capecelatro2015,Willen2019,Hausmann2024b,Xia2024}). If these cases were simulated with the EL point-particle approach with an accurate velocity disturbance correction, incorrect results would be obtained. 

\subsection{A sphere falling in a fluid at higher Reynolds numbers}
In order to test the novel force correlation that we propose for finite $\mathrm{Re_p}$, the study of the falling particle discussed in the previous section is extended to larger $\mathrm{Re_p}$. Similarly to the configuration of the falling particle in the Stokes regime, the domain size remains $L_x\times L_y\times L_z = 100d_{\mathrm{p}}\times 100d_{\mathrm{p}}\times 100d_{\mathrm{p}}$ for this validation. The density ratio is kept at $\rho_\mathrm{p}/\rho_\mathrm{f}=2000$, and the particle Reynolds number is varied by varying the fluid viscosity. At finite $\mathrm{Re_p}$, the results cannot be compared to an analytical solution, but for a single particle, a reference solution can be obtained with a one-way coupled simulation. In the one-way coupled simulation, the fluid velocity remains zero at all times and the classical Schiller-Naumann correlation, as given in Equation \eqref{eq:CDSchillerNaumann}, provides an accurate drag force on the particle. In addition to the one-way coupled simulations, we perform two two-way coupled Euler-Lagrange point-particle simulations of this case by:
\begin{enumerate}
    \item Solving the volume-filtered NSE with the classical Schiller-Naumann drag correlation, and
    \item Solving the volume-filtered NSE with the newly proposed filtered version of the Schiller-Naumann drag correlation using Equation~ \eqref{eq:correlationforurel}. \end{enumerate}

The simulations are performed with three different particle Reynolds numbers, $\mathrm{Re_p}\in [1.11\times 10^{-4},0.9646,38.7]$ and two relative filter widths $\sigma^\prime\in[1,4]$, whereas $\sigma/\Delta x=1$ for all simulations. 

\begin{figure}
    \centering
    \includegraphics[width=\textwidth]{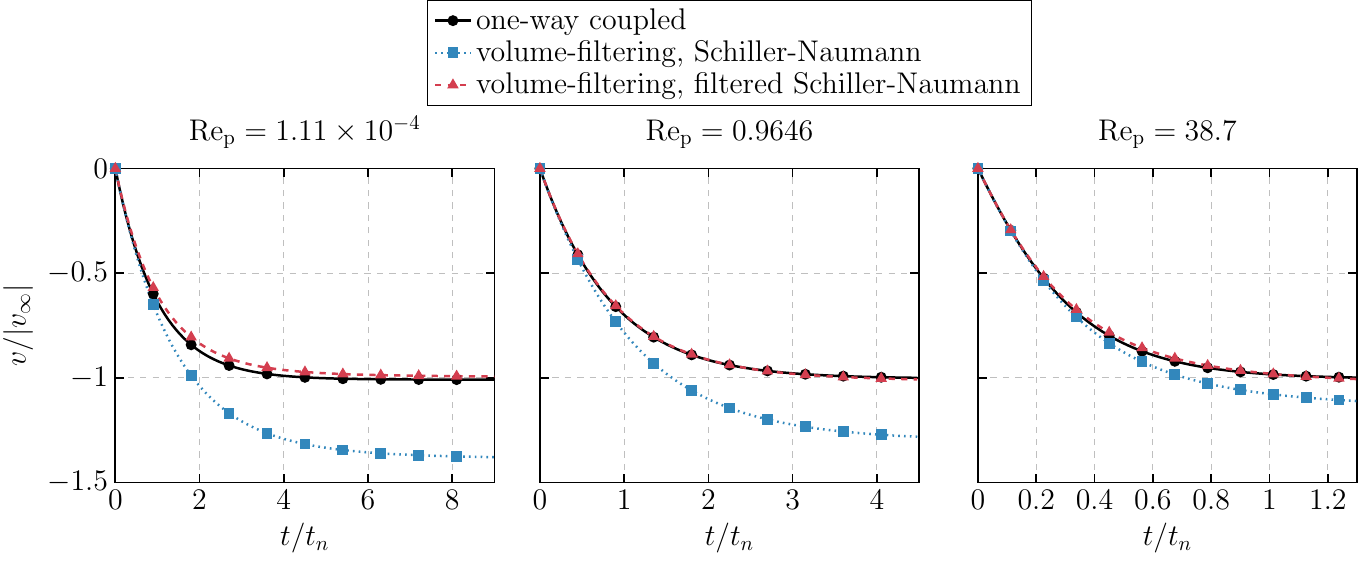}
    \caption{Relative particle velocities of a single falling sphere over time for the volume-filtered simulations using classical Schiller-Naumann drag and the volume-filtered simulations with the filtered Schiller-Naumann drag. The corresponding one-way coupled simulation with classical Schiller-Naumann drag is shown as reference. The simulations are performed with three different values for $\mathrm{Re_p}$ and with a relative filter width of $\sigma^\prime=1$}
    \label{fig:sedimentingSchillerNaumannsigma1}
\end{figure}

\begin{figure}
    \centering
    \includegraphics[width=\textwidth]{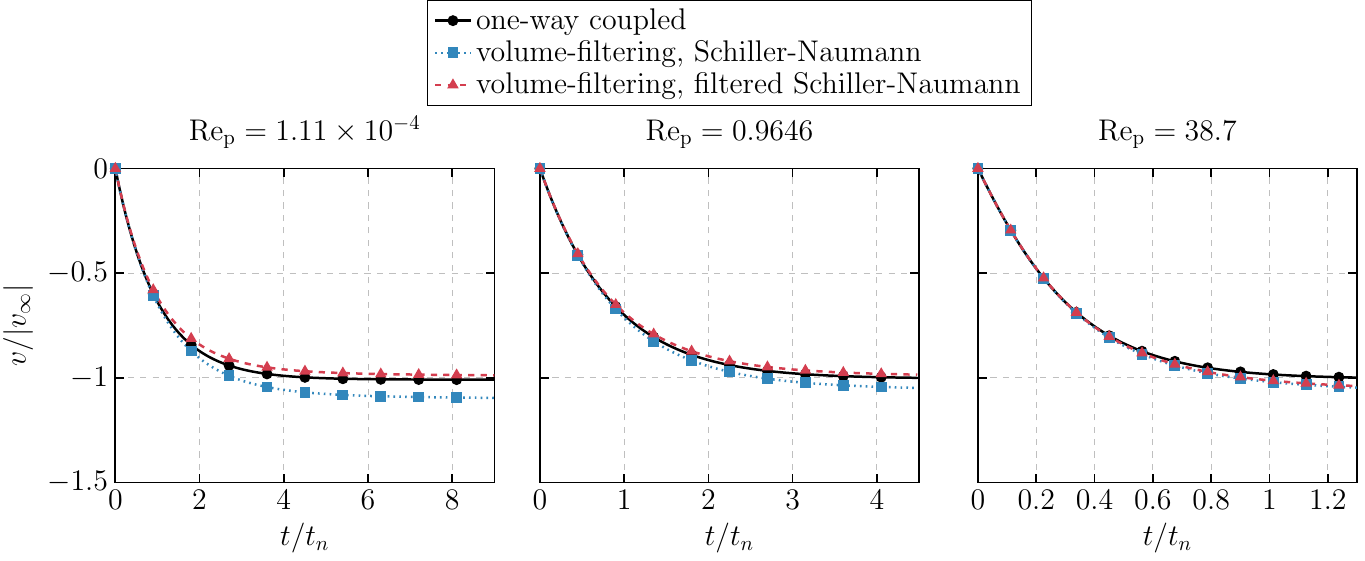}
    \caption{Particle velocities of a single falling sphere over time for the volume-filtered simulations using classical Schiller-Naumann drag and the volume-filtered simulations with the filtered Schiller-Naumann drag. The corresponding one-way coupled simulation with classical Schiller-Naumann drag is shown as reference. The simulations are performed for three different values of $\mathrm{Re_p}$ and with a relative filter width of $\sigma^\prime=4$}
    \label{fig:sedimentingSchillerNaumannsigma4}
\end{figure}

In figure~\ref{fig:sedimentingSchillerNaumannsigma1}, the particle settling velocities are shown for the three $\mathrm{Re_p}$ and for a relative filter width of $\sigma^\prime=1$. The volume-filtered simulations with the newly proposed filtered Schiller-Naumann drag force model predict the particle settling velocities accurately for all of the three $\mathrm{Re_p}$. With the classical Schiller-Naumann drag, significant deviations from the correct particle settling velocities are observed, whereby the deviations decrease with increasing particle Reynolds number. For large particle Reynolds numbers, the flow disturbance induced by the particle becomes small, and the volume-filtered velocity at the particle position is relatively close to the undisturbed velocity. \\
Figure \ref{fig:sedimentingSchillerNaumannsigma4} shows the particle settling velocities for a larger filter width, $\sigma^\prime=4$. Ideally, the predicted particle settling velocities should not depend on the filter width or the spatial resolution and the same results should be obtained as with $\sigma^\prime=1$. This is the case for the volume-filtered simulations using the filtered Schiller-Naumann drag. With the classical Schiller-Naumann drag, the particle settling velocity is still inaccurate for the smaller $\mathrm{Re_p}$, but for the larger $\mathrm{Re_p}$ the particle settling velocities are predicted accurately. 

This means that, when using the classical Schiller-Naumann drag with the volume-filtering framework, the results deteriorate as the relative spatial resolution increases or the local $\mathrm{Re_p}$ decreases. For large filter widths, the flow disturbance by the particle is spread over a wider region, which leads to a volume-filtered velocity at the particle position closer to the undisturbed velocity than for small filter widths. Therefore, the disturbance is negligible for the largest $\mathrm{Re_p}$ and $\sigma^\prime \geq 4$. 

Although accurate particle settling velocities are predicted with the volume-filtering framework and the newly proposed filtered Schiller-Naumann drag, some small deviations to the one-way coupled simulations remain, which have multiple origins: (\romannumeral 1) The volume-filtered NSE are solved on a discrete fluid mesh, which leads to a discretization error. Furthermore, the volume fraction, the mass source, and the closures, including the analytical viscous closure, are also represented on a discrete fluid mesh. The resulting discretization errors lead to a deviation of the numerical solution from the actual volume-filtered flow field, which has been used to fit the coefficients of the force correlation, (\romannumeral 2) The model for the subfilter stress tensor is accurate for small filter widths but the modelling error increases as the filter width increases \cite{Hausmann2024a}, and (\romannumeral 3) A falling particle is a transient process but the filtered Schiller-Naumann correlation is obtained with data from the stationary flow around a sphere. In the acceleration phase, the flow field around the particle can be different from the steady state flow field at the same $\mathrm{Re_p}$. These and other error sources are common in EL point-particle simulations, but the present results indicate that they are small with the volume-filtering framework, at least for the falling sphere configurations investigated. 

\subsection{Suspension of monodisperse spheres}
\begin{figure}
    \centering
    \includegraphics[width=\linewidth]{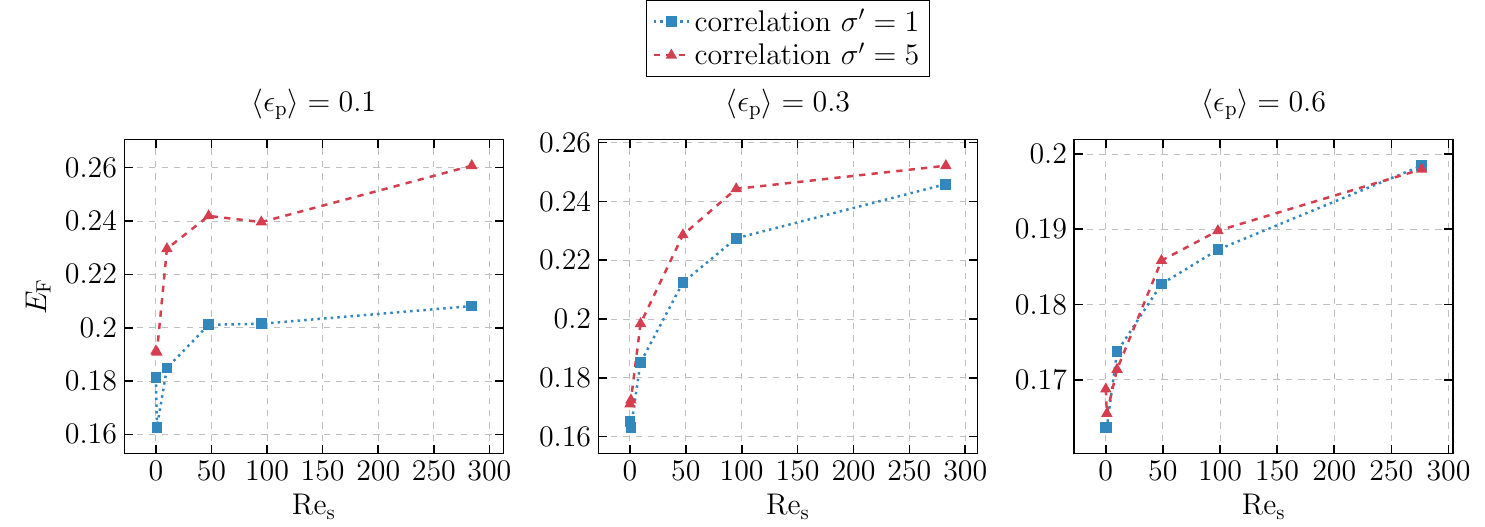}
    \caption{Mean relative force error as a function of the superficial Reynolds number for different filter widths and different global particle volume fractions. }
    \label{fig:errorassempliesRe}
\end{figure}

\begin{figure}
    \centering
    \includegraphics[width=\linewidth]{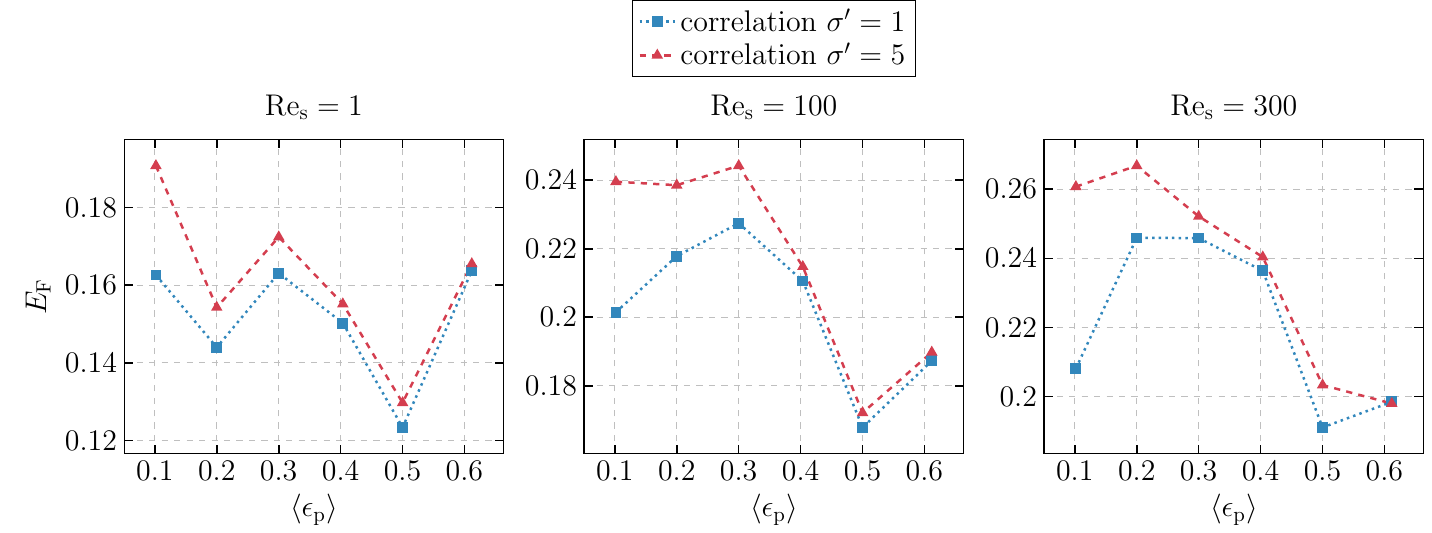}
    \caption{Mean relative force error as a function of the global particle volume fraction for different filter widths and different superficial Reynolds numbers. }
    \label{fig:errorassempliesvfrac}
\end{figure}

As a final test case, the novel force correlation framework is evaluated for a suspension of monodisperse spheres. The hydrodynamical force on each individual particle predicted by the force correlation derived in section \ref{ssec:correlationsuspension} is compared to the actual force given in the PR-DNS of the flow through the particle arrangements. 
In figures \ref{fig:errorassempliesRe} and \ref{fig:errorassempliesvfrac}, the mean relative force error, $E_\mathrm{F}$, of the correlation given in Equation \eqref{eq:filtered_dragcorrelation} is shown for different filter widths, different global particle volume fractions, $\langle \varepsilon_\mathrm{p} \rangle$, and different superficial particle Reynolds numbers, $\mathrm{Re}_\mathrm{s}$.
It should be noted that $\mathrm{Re}_\mathrm{s}$ and $\langle \epsilon_\mathrm{p} \rangle$ are not the filtered quantities at the particle positions but the averages over the entire domain of the respective case, which corresponds to filtered quantities with an infinite filter width. Therefore, the values on the $x$-axes are not the direct values used in the correlation, but they characterize the simulation case, which is why some values on the $x$-axis are the same, although the filter width is different. This means that the mean relative error $E_\mathrm{F}$ is the error observed for a specific superficial Reynolds number and a global particle volume fraction, which is somewhat arbitrary. It may very well be that particles belonging to cases with different superficial velocities and different global particle volume fractions have similar volume-filtered velocities or local volume fractions at the particle positions. \\
The relative errors observed in figures \ref{fig:errorassempliesRe} and \ref{fig:errorassempliesvfrac} are of the order of $20\%$. The proposed correlation cannot be expected to be much more accurate because the volume-filtered velocity and volume fraction are not sufficient to predict the force in such a complex flow with high accuracy. Furthermore, the functional approach used is designed to accurately predict the mean force and not the force on each individual particle. However, more relevant for the proposed framework for force correlations than the magnitude of the error is how the error changes with the filter width. \\
With the filter width $\sigma^\prime=5$, the volume-filtered velocity and volume fraction at the particle positions are almost identical to the superficial velocity and the global particle volume fraction for all particles. Therefore, the correlations of all particles in the same simulation case have almost identical input values, which leads to predicted forces by the correlation that are also almost identical for all particles. For such large filter widths and if only the volume-filtered velocity and the volume fraction are considered as input parameters, the proposed framework yields a correlation that is similar to existing mean force correlations. However, the proposed framework is also suitable for smaller filter widths. As observed in figures \ref{fig:errorassempliesRe} and \ref{fig:errorassempliesvfrac}, the mean relative force error decreases for almost every $\mathrm{Re}_\mathrm{s}$ and $\langle \epsilon_\mathrm{p}\rangle$ as the filter width decreases to $\sigma^\prime=1$. At this filter width, the volume-filtered velocity and volume fraction at the different particle positions varies and the correlation can predict the different drag forces on the particles. To improve the overall accuracy of the force correlation, additional volume-filtered flow quantities need to be included in the correlation, such as the volume-filtered velocity gradient or the volume fraction gradient, which becomes particularly evident for large global particle volume fractions.

As already discussed for the configurations with isolated particles, the proposed framework of drag force correlations does not require the determination of the undisturbed velocity, but is fully based on filtered flow quantities. Moreover, the accuracy of the force prediction does not deteriorate as the filter width or mesh spacing is decreased, as with existing drag force prediction frameworks, but the accuracy of the drag force prediction is improved. 

\section{Conclusions}
This study presents a novel framework for modelling the hydrodynamic forces in Euler–Lagrange (EL) point-particle simulations that does not rely on the classical concept of the undisturbed fluid velocity. By leveraging the volume-filtered Navier–Stokes equations, the proposed approach directly relates the drag force on each of the particles to volume-filtered fluid quantities which are readily available in EL simulations.

The accuracy of the framework has been demonstrated through validation against analytical solutions, one-way coupled simulations, and particle-resolved direct numerical simulations (PR-DNS) across a wide range of particle Reynolds numbers and filter widths. In particular, the filtered drag force correlations accurately recover the correct terminal velocity for a falling sphere in both the Stokes and finite Reynolds number regimes.

Moreover, the framework has been extended to particle assemblies, where a generalized drag correlation has been derived based on the volume-filtered velocity and the volume fraction. 
The new correlation shows an improved predictive performance over existing mean-force models, especially at smaller filter widths and mesh resolutions, and at lower volume fractions. 
To increase the accuracy of the newly proposed framework, apart from the volume-filtered fluid velocity, additional volume-filtered quantities, such as the volume-filtered fluid velocity gradients, are required.

Unlike traditional approaches that depend on estimating undisturbed velocities, the proposed methodology remains robust in both dilute and moderately dense regimes, avoids high computational cost, and enables the formulation of force models that are inherently consistent with the volume-filtered flow description.
The proposed framework provides a theoretically sound and computationally efficient basis for the development of next-generation force correlations in EL point-particle simulations of multiphase flows.

\begin{acknowledgments}
\textbf{Funding.}
This research was funded by the Deutsche Forschungsgemeinschaft (DFG, German Research Foundation): \\
\textemdash Project-ID 457515061, \\
\textemdash Project-ID 457509672, \\
\textemdash Project-ID 422037413 - TRR 287, and\\
\textemdash Project-ID 466092867 - SPP 2331.\\

\textbf{Declaration of interests.} The authors report no conflict of interest. \\

\textbf{Data availability statement.}  
The data that support the findings of this study are reproducible and are openly available in the repository \url{https://doi.org/10.5281/zenodo.15364320}\\

\textbf{Author ORCIDs.}\\
Berend van Wachem: \url{https://orcid.org/0000-0002-5399-4075}\\
Hani Elmestikawy: 
\url{https://orcid.org/0000-0003-0083-2360}\\
Akshay Chandran: \url{https://orcid.org/0000-0002-2273-5601} \\
Max Hausmann: \url{https://orcid.org/0000-0002-4342-4749}

\end{acknowledgments}

\bibliographystyle{jfmc}


\end{document}